\documentclass{eptcs}

\providecommand{\event}{RTRTS 2010} 

\usepackage{breakurl}
\usepackage{graphicx}
\usepackage{latexsym}

\newcommand {\tab} {\mbox{\hspace {0.2in}}}

\title{\textsc{Dist-Orc}: A Rewriting-based Distributed Implementation of Orc with Formal Analysis}
\author{Musab AlTurki
   \institute{The University of Illinois at Urbana-Champaign,\\ 
              Urbana IL 61801, USA}
   \email{alturki@cs.illinois.edu}   
\and Jos{\'e} Meseguer
   \institute{The University of Illinois at Urbana-Champaign,\\ 
              Urbana IL 61801, USA}
   \email{meseguer@cs.illinois.edu}
}

\def\titlerunning{Dist-Orc: A Rewriting-based Distributed Implementation of Orc with Formal Analysis}  
\def\authorrunning{AlTurki and Meseguer}   

\begin{document}

\maketitle              

\begin{abstract}

Orc is a theory of orchestration of services that allows structured programming of  distributed and timed computations. Several formal semantics have been proposed for Orc, including a rewriting logic semantics developed by the authors. Orc also has a fully fledged implementation in Java with functional programming features. However, as with descriptions of most distributed languages, there exists a fairly substantial gap between Orc's formal semantics and its implementation, in that: (i) programs in Orc are not easily deployable in a distributed implementation just by using Orc's formal semantics, and (ii) they are not readily formally analyzable at the level of a distributed Orc implementation. In this work, we overcome problems (i) and (ii) for Orc. Specifically, we describe an implementation technique based on rewriting logic and Maude that narrows this gap considerably. The enabling feature of this technique is Maude's support for external objects through TCP sockets. We describe how sockets are used to implement Orc site calls and returns, and to provide real-time timing information to Orc expressions and sites. We then show how Orc programs in the resulting distributed implementation can be formally analyzed at a reasonable level of abstraction by defining an abstract model of time and the socket communication infrastructure, and discuss the assumptions under which the analysis can be deemed correct. Finally, the distributed implementation and the formal analysis methodology are illustrated with a case study.

\end{abstract}

\section{Introduction}

Concurrent languages for new application areas such as web services
pose interesting challenges.  The usefulness of such languages is
quite clear, but their correctness, both that of a language
implementation and of specific programs in the language, is crucial
for their safety and security.  In particular, one would like to have,
among other things: (i) a clear, semantics-preserving path from a
language specification to a distributed language implementation; (ii)
furthermore, this distributed implementation  path should come
with \emph{formal  correctness guarantees}; and (iii) it
should be possible to \emph{formally verify} that specific programs
written in such a language, and implemented according to the above
path from a language specification, satisfy appropriate formal
requirements.  However, the distributed nature of such languages makes
tasks (i)--(iii) more challenging than for sequential languages.

This paper addresses all these issues for the Orc programming language
\cite{OrcTheory05}, an elegant and powerful language for orchestration of web
services.  Orc has a well-developed theoretical basis \cite{OrcTheory05,OrcTreeSem05,OrcTraceSem06,WSFM07,Wehrman07tcs} and also a
well-engineered language implementation \cite{KitchinQCM09forte,OrcJavaImp05}.  But as for any other
language, there is a substantial gap between a language's
theoretical description and its implementation.  Ideally, we would
like to reason, and obtain formal guarantees about, Orc programs in
their implemented form, but this is not a trivial matter.

{\bf Our Approach}.  In this paper we propose and demonstrate the
effectiveness of a method to substantially narrow the gap between the
theoretical level of a distributed language like Orc and an actual
implementation.  Our method is \emph{semantics-based} and builds on
our earlier work on a \emph{rewriting logic semantics} of Orc
\cite{AlTurkiM07PPDP}, in which we showed how subtle issues about the
Orc semantics, including its real-time nature, and the essential
priority that internal events in an Orc expression should have over
external communication, could be faithfully modeled in our real-time rewriting
semantics.  It also builds on our subsequent work giving to Orc a more
efficient \emph{reduction semantics}
\cite{AlTurkiM07WWV,AlTurkiM07tr}, which brought the Orc language
definition closer to an actual distributed
implementation while preserving semantic equivalence.  The third step,
taken in this paper, is to pass from a \emph{rewriting-based reduction
  semantics} to a \emph{rewriting-based implementation} of such a
semantics in a seamless way.  Since the original SOS-like semantics of
Orc and the reduction semantics are semantically equivalent
\cite{AlTurkiM07tr}, this substantially narrows the gap between the
language's formal semantics and its implementation.

How is this correct-by-construction implementation accomplished?  The
key idea is that concurrent rewriting is \emph{both} a theoretical
model and a practical method of distributed computation.
Specifically, in the Maude language \cite{maude-book}, a high-performance
implementation of rewriting logic, asynchronous message-passing
between distributed objects is accomplished by
concurrent rewriting via sockets.  For Orc, the objects 
are either Orc expressions, which play the role of clients, or web
sites, which play the role of servers.  But since for Orc real time is
of the essence, an important issue that must be addressed is how real
time is supported in the implementation.  Here, the key observation is
that Orc programs assume an asynchronous and possibly unreliable
distributed environment such as the Internet, and therefore implicitly
rely on their \emph{local} notion of time for their computations.  As
a consequence, time is supported by \emph{local ticker objects}, that
interact in a tightly-coupled way with their co-located Orc
objects.

How about formal verification of Orc programs?  Specifically, how can
we model check the formal requirements of an Orc program running in
the rewriting-based distributed implementation just described?  The
answer is that we cannot model check such a program \emph{directly}
with existing tools, but that we can however model check it
\emph{indirectly}.  The idea is in a sense quite simple, namely, we
can \emph{formally specify} both the internet sockets supporting the
actual distributed implementation, and the ticker objects supporting
the real-time behavior of Orc expressions.  In this way, both
distributed message-passing computation between Orc expression clients
and web sites, and time elapse are faithfully \emph{simulated} in the
formal specification, in which our desired program can then be model
checked.  As we explain in the paper, under reasonable assumptions
about the granularity of time chosen for the tickers and the code of
the Orc expressions, this simulated formal analysis gives us
corresponding guarantees about the actual Orc programs running in the
actual distributed Orc implementation.

{\bf Our Contributions}.  Our \emph{first} contribution is the \emph{correct-by-construction} nature of our distributed Orc implementation, henceforth referred to as \textsc{Dist-Orc}.  This contribution builds on the semantic equivalences already proved in \cite{AlTurkiM07WWV} between the SOS, reduction semantics, and object-based semantics levels.  In this way, a seamless path from formal language definition to a correct distributed implementation is obtained.

Our \emph{second}  contribution is to show for the first
time how a rewriting logic specification of an object-oriented real-time system can be
naturally transformed into an actual \emph{distributed implementation} of such
a system in physical time.  For untimed object-based systems it was
known that Maude sockets could be used for this purpose; but no technique was known
for seamlessly passing from real-time specifications to their implementations.  This is
shown here for \textsc{Dist-Orc}, but the method is much more general and has already been
applied to other real-time systems, like medical devices, in recent work
by Sun and Meseguer~\cite{SunM10}.  

A \emph{third}  contribution is to show that we can still formally
verify correctness properties of a distributed real-time
\emph{implementation}  by modeling the implementation
itself at an appropriately chosen level of abstraction.  This had been done for
untimed systems such as Mobile Maude~\cite{maude-book}, but it is done here for real-time systems for the first time.

The paper is organized as follows. In Section \ref{sec:orc}, we give an overview of the Orc theory and its informal semantics, and describe in some detail the \textsc{Auction} example in Orc. In Section \ref{sec:rewriting-semantics}, we briefly review Orc's rewriting semantics previously developed that form the basis of the distributed implementation \textsc{Dist-Orc}, which is presented and discussed in Section \ref{sec:dist-orc}, along with a distributed implementation \texttt{Dist-Auction} of the auction example. Section \ref{sec:formal-analysis} describes the formal model of \textsc{Dist-Orc} in Real-Time Maude, with which formal analysis of distributed Orc applications, such as \texttt{Dist-Auction}, can be carried out. The paper is concluded with a summary and a brief discussion of future work.

\section{An Overview of Orc} \label{sec:orc}

Orc is a theory of orchestration of services that provides an elegant model for describing concurrent computations. Orc uses the notion of a \emph{site} to represent a general service, which may vastly range in complexity from a simple function to a complex web search engine. A site may also represent the interaction with a human being, most commonly within the context of business workflows~\cite{Aalstetal2003}. A site, when called, produces at most one value. A site may not respond to a call, either by design or as a result of a communication problem. For example, if $\mathit{CNN}$ is a site that returns the news page for a given date $d$, then $\mathit{CNN}(d)$ might not respond because of a network failure or it may choose to remain silent because of an invalid value $d$. When a site responds to a call with a value $c$, it is said to \emph{publish} the value $c$. Site calls are \emph{strict}, in that a site 
call cannot be initiated before its parameters are bound to concrete values.

Orc's computation model is timed. Different site calls may occur at different times. A site call may be purposefully delayed using the \emph{internal} site $\mathit{rtimer}(t)$, which publishes a signal after exactly $t$ time units. Furthermore, responses from calls to \emph{external} sites may experience unpredictable delays and communication failures, which could affect whether and when other site calls are made. Unlike external sites, however, responses from internal sites, such as $\mathit{rtimer}$, are assumed to have completely predictable timed behaviors. Orc also assumes a few more internal sites, which are needed for effective programming. They are: (1) the $\mathit{if}(b)$ site, which publishes a signal if $b$ is true and remains silent otherwise, (2) $let(\bar{x})$, which publishes a tuple of the list of values in $\bar{x}$, and (3) $\mathit{clock}$, which publishes the current time value.

\begin{figure}
	\centering
	\begin{small}
\begin{tabular}{cc}		
	\begin{tabular}{rcl}
	        Orc program  & ::= & $\bar{d} ~;~ f$ \\
	$d \in$ Declaration  & ::= & $E(\bar{x}) =_{def} f$ \\
	$f,g \in$ Expression & ::= & $\mathbf{0}$ $|$ $M(\bar{p})$ $|$ $E(\bar{p})$
						 $|\;\;\;$    $f \;|\; g$ 
						  $\;\;\;|\;\;\;$    $f > x > g$ 
						  $\;\;\;|\;\;\;$    $g < x < f$ \\
	$p \in$ Actual Parameter & ::= & $x \;|\; c \;|\; M$ \\
	\end{tabular}
\end{tabular}
\end{small}
	\caption{Syntax of Orc}
	\label{fig:syntax}
\end{figure}

Orchestration \emph{expressions} in Orc describe how individual site calls (and responses) are combined in order to accomplish a larger, more useful task. Orc expressions are built up from site calls using only three combinators, which were shown to be capable of expressing a wide variety of distributed computations succinctly and elegantly~\cite{OrcTheory05}. 
The syntax of Orc is shown in Figure \ref{fig:syntax}. An Orc \emph{program} consists of an optional list of declarations, giving names to expressions, followed 
by an Orc expression to be evaluated. 
An \emph{expression} can be either: (1) the silent expression ($\mathbf{0}$), 
which represents a site that never responds; (2) a site or an expression call having 
an optional list of actual parameters ($M(\bar{p})$ and $E(\bar{p})$, respectively); or (3) the composition of two expressions by one of the three composition operators: 

\textbf{Symmetric parallel composition}, $f \;|\; g$, models concurrent execution of independent threads of computation. For example, $\mathit{CNN}(d) \;|\; \mathit{BBC}(d)$, where $\mathit{CNN}$ and $\mathit{BBC}$ are sites, calls both sites concurrently and may publish up to two values depending on the publication behavior of the individual sites.

\textbf{Sequential composition}, $f > x > g$, executes $f$, and for every value $c$ published by $f$, a fresh instance of $g$, with $x$ bound to $c$, is created  and run in parallel with $f > x > g$. This generalizes sequencing expressions in traditional programming languages, where $f$ publishes exactly one value. For example, if $\mathit{Email}(a, x)$ is a site that sends an e-mail message given by $x$ to a fixed address $a$, then the expression $\mathit{CNN}(d) > x > \mathit{Email}(a, x)$ may cause a news page to be sent to $a$. If $\mathit{CNN}(d)$  does not publish a value, $\mathit{Email}(a, x)$ is never invoked. Similarly, the expression $(\mathit{CNN}(d) \;|\; \mathit{BBC}(d)) > x > \mathit{Email}(a, x)$ may result in sending zero, one, or two messages to $a$. 

\textbf{Asymmetric parallel composition}, $f < x < g$, executes $f$ and $g$ concurrently but terminates $g$ once it has published its first value, which is then bound to $x$ in $f$. For instance, the expression $\mathit{Email} (a, x) < x < (\mathit{CNN}(d) \;|\; \mathit{BBC}(d))$ sends at most one message to $a$, depending on which site publishes a value first. If neither site publishes, the variable $x$ is not bound to a concrete value, and the call to $Email$ is not made.


To minimize use of parentheses, we assume that sequential composition has precedence over symmetric parallel composition, which has precedence over asymmetric composition. We also use the syntactic sugar $f \gg g$ for sequential composition when no value passing from $f$ to $g$ is taking place, which corresponds to the case of a sequential composition $f > x > g$ where $x$ is \emph{not a free} variable of $g$. For example, the Orc expression $\mathit{let}(x) < x < (\, M \;|\; \mathit{rtimer}(t) \gg let(0)\, )$ specifies a timeout on the call to a site M. 
Upon executing it, both sites $M$ and $rtimer$ are called. If $M$ publishes a value $c$ before $t$ time units, then $c$ is the value published by the expression. But if $M$ publishes $c$ in exactly $t$ time units, then either $c$ or $0$ is published. Otherwise, $0$ is published.
Another example is the two-branch conditional $\mathbf{if}\;b\;\mathbf{then}\;f\;\mathbf{else}\;g$, which can be succinctly specified in Orc as the expression $\mathit{if}(b) \gg f \;\;|\;\; \mathit{if}(\neg b) \gg g$. 
Given the behavior of the internal site $\mathit{if}$, only one of $f$ and $g$ is executed, depending on the truth value of $b$. 
These examples and many more can be found in~\cite{OrcTheory05}.

\textbf{The Auction Example}. This section concludes with a description of a larger example in Orc, namely \textsc{Auction}, which will be the basis for a case study illustrating \textsc{Dist-Orc} in Section \ref{sec:dist-orc}, and its formal analysis in Section \ref{sec:formal-analysis}.
The Orc program \textsc{Auction}, which was originally inspired by the auction example in~\cite{OrcTheory05}, is a simplified online auction management application that manages posting new items for auction, coordinates the bidding process, and announces winners. 

\textsc{Auction} assumes the following sites: (1) A $\mathit{Seller}$ site, which maintains a list of items to be auctioned and responds to the message \textsf{postNext} by publishing the next available item, (2) a $\mathit{Bidders}$ site, which maintains a list of bidders and their bids, and responds to the message \textsf{nextBidList},  which solicits a list of bids higher than the current bid for the auctioned item, (3) a $\mathit{MaxBid}$ site, which is a functional site that publishes the highest bid of a list of bids, and (4) an $\mathit{Auction}$ site, which maintains a list of available items and responds to \textsf{post} and \textsf{getNext} for adding and retrieving an item from the list, respectively. The $\mathit{Auction}$ site also keeps a list of winners and the respective items won, and responds to the message \textsf{won}, which declares a bidder as the winner for the auctioned item.

\begin{figure} 	
	\centering
\begin{small}
\[
\begin{array}{l}
\mathit{Posting}(\mathit{seller}) 
    =_{def} \mathit{seller}(\mathsf{``postNext"}) > x > \mathit{Auction}(\mathsf{``post"}, x) \gg 
     \mathit{rtimer}(1) \gg \mathit{Posting}(\mathit{seller}) 
    \vspace{0.07in} \\
    
\mathit{Bidding} 
   =_{def}  \mathit{Auction}(``\mathsf{getNext}") > (id, d, m) > 
        \mathit{Bids}(id, d, m, 0) > (wn, wb) > \\
    \tab\tab\tab ( \; \mathit{if} (wn = 0) \gg \mathit{Bidding}() \\
    \tab\tab\tab \; | \; \mathit{if} (wn \neq 0) \gg \mathit{Auction}(``\mathsf{won}", wn, id, wb) \gg \mathit{Bidding}() \;\;) 
        \vspace{0.07in} \\

\mathit{Bids} (id, d, wb, wn) 
   =_{def}  (\;  \mathit{if} (d \leq 0) \gg \mathit{let}(wb, wn) \\
    \tab\tab\tab       | \; \mathit{if} (d > 0) \gg \mathit{clock}() > t_{a} > \mathit{min}(d, 1) > t > \mathit{TimeoutRound}(id, wb, t) > x > \\
     \tab\tab\tab\tab (\; \mathit{if}(x = \mathit{signal}) \gg \mathit{Bids}(id, d - t, wb, wn) \\
     \tab\tab\tab\tab | \; \mathit{if}(x \neq \mathit{signal}) \gg    
                   \mathit{rtimer}(1) \gg \mathit{clock}() > t_{b} > \mathit{Bids}(id, d - (t_{b} - t_{a}), x_{0}, x_{1}) 
            \; )
        \; ) 
            \vspace{0.07in} \\

\mathit{TimeoutRound} (id, bid, t) 
   =_{def}  \\ 
   \tab\tab\tab let(x) < x < ( \; \mathit{rtimer}(t) \;\; | \;\; \mathit{Bidders}(``\mathsf{nextBidList}", id, bid) > bl > \mathit{MaxBid}(bl) \;) 
\end{array}
\]
\end{small}
	\caption{Orc expressions in the \textsc{Auction} program}
    \label{fig:Auction-exprs}
\end{figure}

Figure \ref{fig:Auction-exprs} lists Orc expressions used by \textsc{Auction}. 
The $\mathit{Posting}$ expression recursively queries a given seller site for the next item available for auction $x$, and then posts it to the auction by calling the $\mathit{Auction}$ site. The call to $\mathit{Auction}$ blocks until bidding on $x$ has ended. The $\mathit{Posting}$ expression then waits for one more time unit before querying the seller for the next item.

The $\mathit{Bidding}$ expression recursively queries the auction site for the next item available for auction and collects bids for the item in rounds from the bidders site for the duration of the auction, where each round lasts for a maximum of one unit of time. Once the bidding ends, the $\mathit{Bidding}$ expression then announces the winning bidder before proceeding to the next item. An item is a tuple $(id, d, m)$, with $id$ the item identifier, $d$ its auction duration, and $m$ the starting bid. We use integer subscripts in variables to pick elements of a tuple, e.g. $x_{0}$ is the first element of the tuple given by $x$, and so on.

The declarations in Figure \ref{fig:Auction-exprs} along with the expression $\mathit{Posting(s)} \;|\; \mathit{Bidding}$, for a given seller site $s$, fully specify in Orc the $\mathit{Auction}$ program:
\[ \mathit{Posting}\;;\; \mathit{Bidding}\;;\; \mathit{Bids}\;;\; \mathit{TimeoutRound}\;;\;\;  \mathit{Posting(s)} \;|\; \mathit{Bidding}\]

\section{Rewriting Semantics of Orc} \label{sec:rewriting-semantics}

Rewriting logic \cite{RL92} 
is a general semantic framework that is well suited for giving formal  
definitions of programming languages and systems, including their concurrent and real-time features (see
\cite{meseguer-rosu-ijcar04,meseguer-rosu-tcs,journ-rtm} and references there).
Furthermore, with the availability of high-performance
rewriting logic implementations, such as Maude~\cite{maude-book},
language specifications can both be executed and model checked.

In previous work~\cite{AlTurkiM07WWV}, we have developed an executable specification giving a formal semantics to Orc in rewriting logic. The specification was shown to capture Orc's intended \emph{synchronous} semantics~\cite{AlTurkiM07tr}, where actions internal to an Orc expression, namely site calls, expression calls, and publishing of values, are given priority over interactions with the environment (processing responses from external sites), while also capturing its timed behaviors. Furthermore, our specification went through three main semantics-preserving refinements~\cite{AlTurkiM07PPDP,AlTurkiM07tr,AlTurkiM07WWV} in order to achieve greater efficiency and expressiveness. 

The distributed implementation presented in this paper builds on the object-based semantics $\mathcal{R}_{Orc}$, which was first introduced in \cite{AlTurkiM07WWV}. This semantics generalizes the previous semantics to multiple Orc expressions and makes explicit the interactions between Orc site and expression objects. 
The extension to object-based semantics enabled defining arbitrarily complex applications in Orc. We give a quick overview of the object-based semantics below.

In $\mathcal{R}_{Orc}$, the state of an Orc program is defined as a \emph{configuration} (of sort $\mathsf{Configuration}$) of objects and messages. Configurations are multisets specified by the associative and commutative empty juxtaposition operator $\_\_ : \mathsf{Configuration} \; \mathsf{Configuration} \to \mathsf{Configuration}$, with the empty multiset $\mathit{null}$ as the identity element. An object is a term of the form $\langle I : \mathcal{C}\;|\; \mathit{AS} \rangle$, with $I$ a unique object identifier, $\mathcal{C}$ the class name of the object, and $\mathit{AS}$ a set of attribute-value pairs, each of the form $\mathit{attr} : \mathit{value}$, where $\mathit{attr}$ is the attribute's name, and $\mathit{value}$ is the corresponding value. There are two main classes of Orc objects: \emph{Expression} objects and \emph{Site} objects, corresponding, respectively, to Orc expressions and sites.  An Orc expression object for an Orc program $\bar{d} ~;~ f$ maintains a set of expression declarations corresponding to $\bar{d}$ in an attribute $\mathit{env}$, and the expression $f$ to be evaluated in an attribute $\mathit{exp}$. An Orc site object has a \emph{name} attribute for the site's name, and maintains the site's state in a $\mathit{state}$ attribute.

In keeping with the philosophy of the Orc theory, expression objects are modeled as active objects with one or more threads of control (given as an Orc expression), and are capable 
of initiating (asynchronous) message exchange. Site objects, on the other hand, are reactive objects having internal states but are only capable of responding to incoming requests. In order to capture timing behaviors, an additional simple \emph{Clock} object is also included in the configuration.

Messages in an Orc configuration are either site call or site return messages. Within an expression object $O$, a site call expression $M(C)$ causes a site call message of the form $M \gets sc(O, C, R)$ to be emitted into the configuration, with $R$ a non-negative value representing the delay of the message; that is, the time it takes for the message to reach the site $M$. Once the message is received and processed by $M$, the site may reply back with a site return message $O \gets sr(M,c, R')$, with $c$ the value published by $M$, and $R'$ a message transmission delay.

Timed behaviors are specified using a \emph{tick} rule that advances logical time and manages the effect of time elapse by appropriately updating message delays. To provide proper timing guarantees for internal sites and to rule out uninteresting behaviors, a rule application strategy that gives time ticking the lowest priority among all rules in the specification is used. For the analysis to be mechanizable, we also assume Orc programs with ``non-Zeno'' behaviors~\cite{OlveczkyM2007}, such that only a finite number of instantaneous transitions are possible within any finite  period of time~\cite{AlTurkiM07WWV}.

In the rewriting semantics of Orc outlined above, although many of the concepts and techniques related to specifying timed behaviors in rewrite theories were borrowed from work on \emph{real-time rewrite theories}~\cite{journ-rtm} and their implementations in Real-Time Maude~\cite{RTMManual07}, the tools and infrastructure provided by Core Maude were enough for our modeling and analysis purposes. However, in Section \ref{sec:formal-analysis}, we use Real-Time Maude to arrive at a more flexible and expressive formal model for the distributed implementation presented in the next section, and use its formal analysis tools, such as time-bounded model-checking.

\section{A Distributed Implementation of Orc} \label{sec:dist-orc}

The Orc theory was designed to specify, in a structured manner, concurrent computations, with emphasis on distribution through the notion of (external) sites. Furthermore, in practical applications it is typically the case that an Orc expression combines (through the use of Orc combinators) several subexpressions that independently orchestrate different but related tasks. For example, an online auction management expression may be composed of subexpressions managing: (1) seller inventories and product auction announcements, (2) bid collection and winner announcements, and (3) payments and shipping coordination. Such subexpressions need not be located on the same machine for the orchestration effort to be completed, but are, in fact, more often run on physically distributed autonomous agents spread across the web. We, therefore, describe an extension of the Orc theory to a distributed programming model that is both natural and useful in specifying and analyzing distributed computations with explicit treatment of external sites and messages. The extension encapsulates the Orc programming model as the underlying model for Orc expressions, and in this respect, its rewriting specification builds on the Orc rewriting semantics specification $\mathcal{R}_{Orc}$ 
outlined in Section~\ref{sec:rewriting-semantics}. 

In general, the method of transforming a real-time, object-based rewriting semantics into a real-time distributed implementation consists of three fundamental steps:
\begin{enumerate}
\item Defining the distributed structure of the system being specified by specifying locations and a globally unique naming mechanism for objects.
\item Specifying the rewriting semantics of the underlying communication model for distributed objects in the system.
\item Devising a mechanism for capturing real, wall clock timing information and extending the rewriting semantics of time to incorporate this information.
\end{enumerate}
As explained below, a crucial enabling feature for steps (2) and (3) above is Maude's support for socket-based communication~\cite{maude-book}. We give a brief overview of Maude's implementation of sockets below. We then discuss in some detail how this method is applied to Orc's rewriting semantics, outline the design and implementation choices in \textsc{Dist-Orc} and explain how they are specified in Maude. The full specification is available online at \texttt{\url{http://www.cs.illinois.edu/homes/alturki/dist-orc}}.

\subsection{Sockets and External Objects in Maude}

Maude provides a low-level implementation of sockets, which effectively enables a Maude process to exchange messages with other processes, including other Maude instances, according to the connection-oriented TCP communication protocol. More specifically, a configuration \texttt{CF} that contains a \emph{socket portal}, which is a predefined constant \texttt{<>} of sort \texttt{Portal} (which is a subsort of \texttt{Configuration}), may communicate with objects external to the Maude process executing \texttt{CF} through a set of special messages defining an interface to Maude sockets. Assuming that \texttt{O} is an identifier for an object in \texttt{CF}, these messages are as follows:

\begin{enumerate}
\item \texttt{createClientTcpSocket(socketManager, O, ADDRESS, PORT)}, which asks Maude's socket manager, a factory for socket objects,  for a client socket to a server located at \texttt{ADDRESS:PORT}. Maude then responds with either a \texttt{createdSocket(O, socketManager, SOCKET)} message, indicating successful creation of the client socket \texttt{SOCKET}, or a \texttt{socketError(O, socketManager, S)} message, with \texttt{S} a string briefly describing the reason for failure.

\item \texttt{send(SOCKET, O, S)}, which asks for the string \texttt{S} to be sent through \texttt{SOCKET}. This message elicits either a message \texttt{sent(O, SOCKET)}, when the string is successfully sent, or a message \texttt{closedSocket (O, SOCKET, S)}, if an error occurred.	

\item \texttt{receive(SOCKET, O)}, which solicits a response through \texttt{SOCKET}. When a response is received, Maude issues the message \texttt{received(O, SOCKET, S)}, with \texttt{S} the string received. In case of an error, the socket is closed with the message \texttt{closedSocket(O, SOCKET, S)}.

\item \texttt{createServerTcpSocket(socketManager, O, PORT, BACKLOG)}, which asks Maude's socket manager to create a server socket at port \texttt{PORT}, with \texttt{BACKLOG} a positive integer specifying the maximum allowed number of queue requests. The message elicits either a message\texttt{createdSocket(O, socketManager, SERVER-SOCKET)}, or a message \texttt{socketError(O, socketManager, S)}.
 
\item \texttt{acceptClient(SERVER-SOCKET, O)}, which causes Maude to listen for incoming connections at \texttt{SERVER-SOCKET}. If a client connection is accepted, Maude responds back with the message \texttt{acceptedClient (O, SERVER-SOCKET, ADDRESS, SOCKET)}, where \texttt{ADDRESS} is the client's address and \texttt{SOCKET} is a newly created socket for communicating with the client. The message \texttt{socketError(O, socketManager, S)} is issued in case of failure.

\item \texttt{closeSocket(SOCKET, O)}, which causes Maude to close the socket and issue the message \texttt{closedSocket(O, SOCKET, S)}.

\end{enumerate}

Rewriting a term with external objects in Maude is performed with the \texttt{erewrite} command (also abbreviated as \texttt{erew}). A more detailed discussion of sockets and support for external objects in Maude can be found in \cite{maude-book}.

\subsection{Distributed Orc Configurations}

In the distributed implementation, an Orc configuration may span multiple nodes in an interconnected network, and is thus called a \emph{distributed} Orc configuration. Both expression and external site objects in a distributed configuration are identified partly by their \emph{location} (a term of the sort \texttt{Loc}), which is defined as a combination of an address (such as a URI or an IP address) and a port number. To fully identify expression and external site objects, expression object identifiers, of sort \texttt{EOid}, and external site identifiers, of sort \texttt{XSOid}, also include a locally unique sequence number:

\begin{small}
\begin{verbatim}
  op loc : String Nat -> Loc .
  op s : Loc Nat -> XSOid .     
  op e : Loc Nat -> EOid .
\end{verbatim}
\end{small}

Internal site objects, such as $\mathit{if}$ and $\mathit{rtimer}$, are identified simply by their names, since their locations are implicit.

Within a distributed Orc configuration, a \emph{local configuration} at some node, or simply a \emph{configuration}, is managed by an independent instance of Maude. In addition to expression and site objects, each such configuration contains a clock object for maintaining local time and a socket portal for exchanging messages with external objects in other configurations. Local configurations are encapsulated by an operator \texttt{ op [\_] :~Configuration -> LocalSystem } to support managing time and its effects. 

\subsection{Sockets and Messaging}\label{sec:sockets-messaging}

In agreement with the Orc theory, the communication model between Orc expressions and sites follows very closely that of the client-server architecture, where Orc expressions are client objects requesting and using services from sites as server objects.
In particular, when an expression object \texttt{O} makes a site call with actual parameters \texttt{C} to an external site identified by \texttt{s(loc(SR, PT), N)}, a site call message of the form  \texttt{s(loc(SR, PT), N) <- sc(O, C, H)} is created within the configuration, where \texttt{H} is a temporary handle that uniquely identifies the call. This message then triggers a request for creating a client socket to the called site through the following equation:

\begin{small}
\begin{verbatim}
eq s(loc(SR, PT), N) <- sc(O, C, H)  
   = < p(O, H) : Proxy | param : C, response : "" >
     createClientTcpSocket(socketManager, p(O, H), SR, PT) .
\end{verbatim}
\end{small}

\noindent In addition, the equation creates a temporary \emph{proxy} object identified by \texttt{p(O, H)}, which manages external communication for this particular site call on behalf of the expression object \texttt{O}. The proxy object also serves as a buffer for the site's response, since TCP sockets do not preserve message boundaries in general.

If a client socket to \texttt{loc(SR, PT)} is successfully created, the message \texttt{ createdSocket(OP, socketManager, SC)}, targeted to the proxy \texttt{OP}, is introduced into the configuration. This causes the proxy to send the site call message to the external site object through the socket \texttt{SC}, as specified by the rewrite rule (the variable \texttt{AS} denotes the rest of the attributes in the object):

\begin{small}
\begin{verbatim}
rl [SendExtCall] :
  createdSocket(OP, socketManager, SC) 
  < OP : Proxy | param: C, AS >
  => < OP : Proxy | param: C, AS > 
     send(SC, OP, (toString(C) + sep)) .
\end{verbatim}
\end{small}

\noindent where \texttt{toString} is a function that properly serializes Orc constants into strings that can be transmitted through sockets. The function uses a separator \texttt{sep} to distinguish message boundaries. 
At the other end, Orc constants are built back from such strings using another function \texttt{toConst}. 

There is also the possibility of an unsuccessful client socket creation attempt due, for example, to an unavailable server or a network failure. In this case, Maude reports the error by issuing the \texttt{socketError(OP, socketManager, S)} message. Such an error is a run-time error, which, for simplicity, is considered fatal in \textsc{Dist-Orc}, so that the site call and any subsequent transitions that depend on it will fail.

Once the site call message is sent, the reply \texttt{sent(OP, SC)} appears in the configuration and the proxy object waits for a response by introducing a \texttt{receive(SC, OP)} message. 

\begin{small}
\begin{verbatim}
rl [RecExtResponse] :
  sent(OP, SC) < OP : Proxy | AS > => < OP : Proxy | AS > receive(SC, OP) .
\end{verbatim}
\end{small}

When some string \texttt{S} is received through the socket, the message \texttt{received(OP, OD, S)} appears, and the proxy object stores \texttt{S} in its buffer and waits for further input. 

\begin{small}
\begin{verbatim}
rl [AccumExtResponse ] :
  received(OP, SC, S) 
  < OP : Proxy | param: C, response: S' >
  => < OP : Proxy | param: C, response: S' + S > 
     receive(SC, OP) .
\end{verbatim}
\end{small}

The proxy will keep accumulating input through the socket until it is remotely closed by the site, when the reply \texttt{closedSocket} appears. The site response is then reconstructed and handed in to its expression object:

\begin{small}
\begin{verbatim}
rl [ProcessExtResponse] :
  closedSocket(p(O,H), SC, S) 
  < p(O,H) : Proxy | response: S' , AS >
  => O <- sr(toConst(S'), H) .
\end{verbatim}
\end{small}

On the server side, when a site is first initialized, it creates a server TCP socket, through which it keeps listening for incoming connections. Once a client has been connected, the site receives the request (which contains the actual parameters for the site call), processes it, and, if appropriate, emits a response back to the client and closes the socket. Site objects employ a similar mechanism for message handling as the one described above.

It is worth noting that, just like any other distributed communication mechanism, messaging through sockets is inherently prone to various potential communication problems, including connection initiation errors, dropped connections, lossy channels and unpredictably long delays. In \textsc{Dist-Orc}, such problems are dynamic errors that might be exposed while executing a distributed Orc program, and typically cause the Orc objects in which they appear to fail.

\subsection{Timed Behavior} \label{sec:timed-implementation}

\begin{figure} 	
	\centering
	\includegraphics[width=0.90\textwidth]{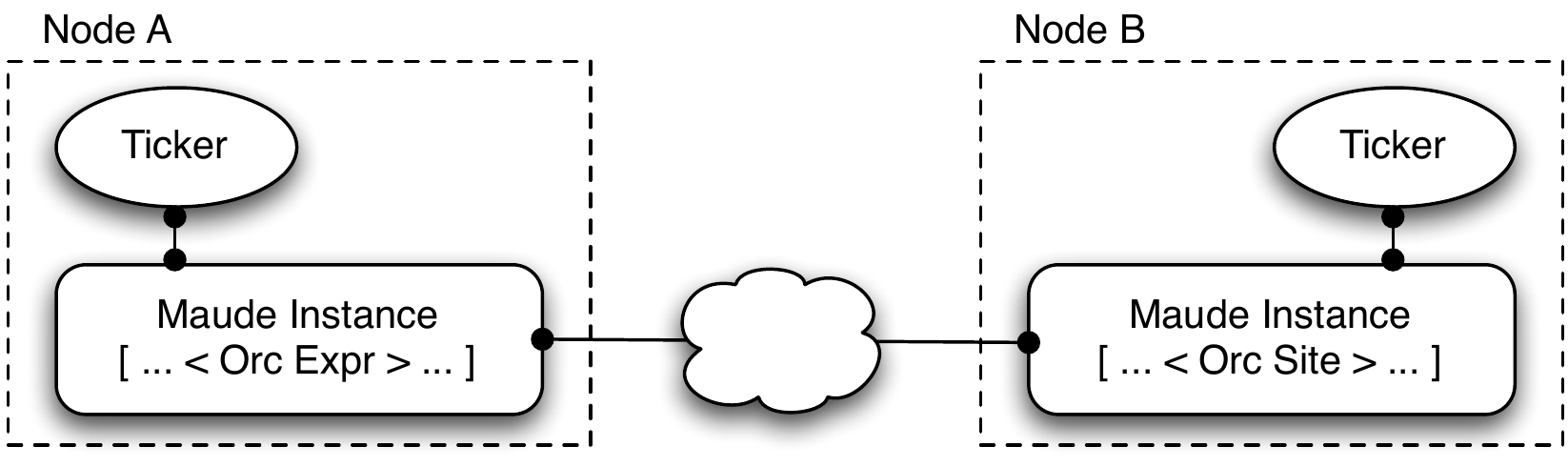}
	\caption{A diagram illustrating the general structure of a distributed Orc configuration. Dashed rectangles represent node boundaries, solid rounded rectangles represent local configurations, and darkened circles represent endpoints of TCP sockets.}
    \label{fig:dist-orc-conf}
\end{figure}

The notion of time in a language implementation is typically captured by a clock against which events in a program in that language may take place. There are several different ways in which clocks can be used to maintain timing information. For our distributed implementation, however, a number of requirements influence the design choices we have made. First, Orc's communication model is asynchronous. This suggests the use of \emph{distributed clocks}, where each node in the distributed configuration maintains its local clock, which also emphasizes Orc's philosophy of having the communicating entities as loosely coupled as possible. Furthermore, for all the applications we have so far specified in Orc, distributed clock synchronization mechanisms, such as Lamport counters \cite{Lamport78} or vector clocks \cite{Mattern89,Fidge88}, are not required for programs correctness. This is primarily due to the fact that in most applications clocking information is used either locally (for example with the local $\mathit{rtimer}$ site) or to time incoming responses. Since in Maude there is no direct support for accessing the system clock, we employ sockets as a means of transmitting clock time ticks to Maude from a \emph{ticker} object external to Maude, but with access to the node's real-time system clock. Thus, for each node in the distributed configuration, the clock of the local Orc configuration is indirectly managed by a corresponding ticker object within the same node. 
It is important to note that, although we use sockets to implement it, the ticker object is local to the node and is thus guaranteed to provide fairly accurate clocking information. Figure \ref{fig:dist-orc-conf} illustrates schematically the deployment architecture of a distributed Orc configuration with timing.

\begin{figure} 	
	\centering
	\includegraphics[width=0.90\textwidth]{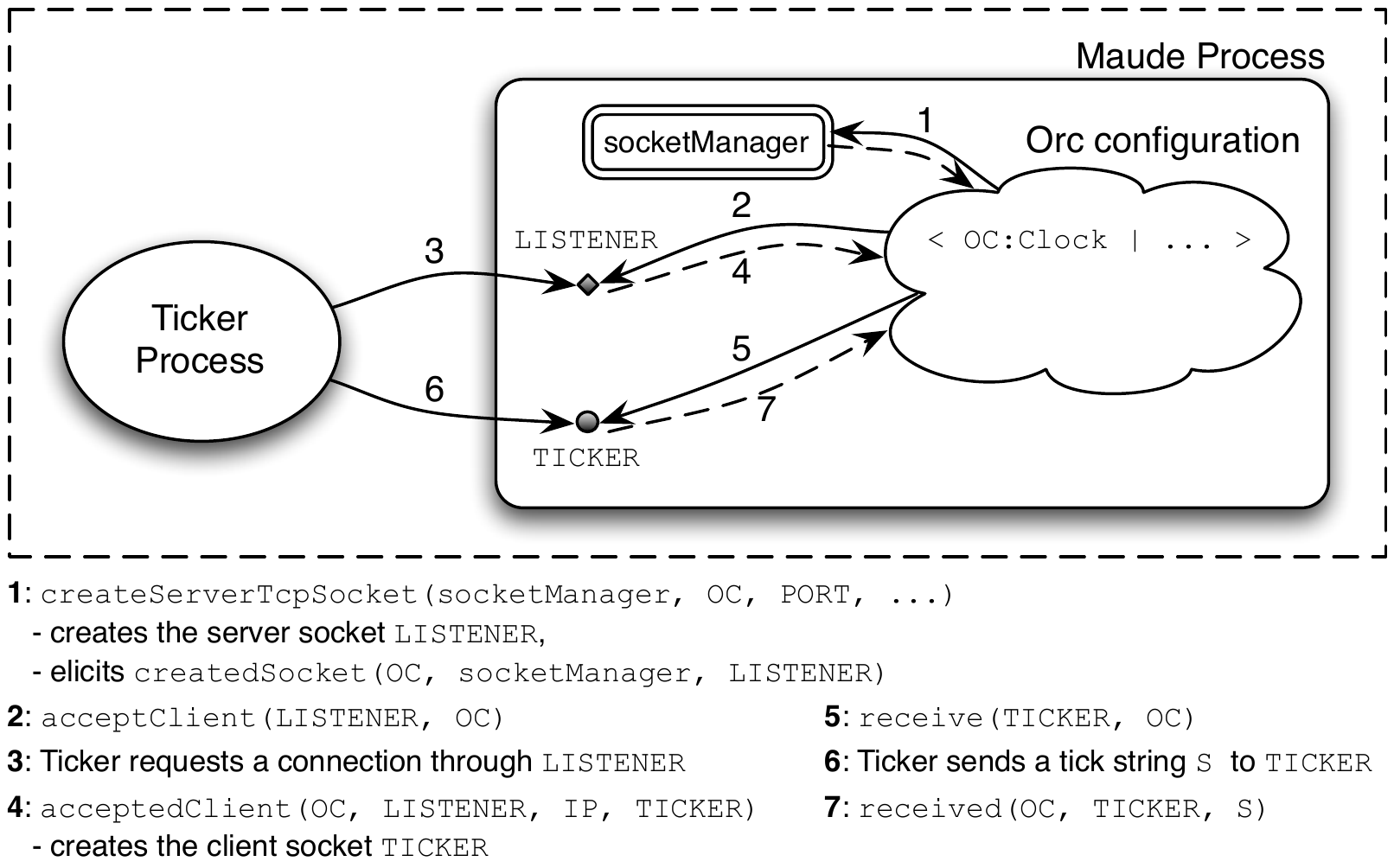}
	\caption{Establishing a connection with the ticker object and receiving clock ticks.}
    \label{fig:ticker-node}
\end{figure}

The diagram in Figure \ref{fig:ticker-node} outlines the steps involved in initializing a connection with the co-located ticker object and receiving the first clock tick.

Upon initialization, the clock object within an Orc configuration requests a server socket, by issuing the message \texttt{createServerTcpSocket(...)}, to be used for listening for a connection from the local ticker process, which is a Java process that is run in every node of a distributed configuration. The ticker process uses the built-in Java classes \texttt{Timer} and \texttt{Socket} to generate and send a tick message every $t$ milliseconds to its corresponding Maude process, where $t$ is a positive integer value. The clock object waits for a connection as soon as the server socket is created:

\begin{small}
\begin{verbatim}
rl [InitClockSocket1] :
  createdSocket(OC, socketManager, LISTENER)
  < OC : Clock | AS >  
  => < OC : Clock | AS > 
     acceptClient(LISTENER, OC) .
\end{verbatim}
\end{small}

\noindent where \texttt{LISTENER} is the newly created clock server socket. 
Once the ticker object is connected, the message \texttt{acceptedClient(OC, LISTENER, IP, TICKER)} appears, with \texttt{IP} the originating address of the ticker object, and \texttt{TICKER} the newly created client socket for communicating with the ticker object. This causes the clock object to become ready for incoming clock ticks according to the following rule:

\begin{small}
\begin{verbatim}
rl [InitClockSocket2] : 
  acceptedClient(OC, LISTENER, IP, TICKER)
  < OC : Clock | AS >  
   => < OC : Clock | AS > 
      receive(TICKER, OC) .
\end{verbatim}
\end{small}

Upon receiving a clock tick, the clock object updates its clock and reflects the effect of time elapse on the rest of the Orc configuration equationally using the time-updating function \texttt{delta}, which decrements the relative time delays in pending messages (see~\cite{journ-rtm} for an explanation of the \emph{delta} methodology). 

\begin{small}
\begin{verbatim}
crl [tick] : 
  [received(OC, TICKER, S) < OC : Clock | clk: c(N) > CF]
   => [< OC : Clock | clk: c(N + 1) > receive(TICKER, OC) delta(CF)]  if ...
\end{verbatim}
\end{small}

\noindent The variable \texttt{CF} denotes the rest of the local configuration. Recall that the operator \texttt{[\_]} encapsulates the local configuration. The process of receiving and processing time ticks repeats as long as the ticker object is supplying ticks through the clock socket.

An important observation is that the use of real, wall clock time in \textsc{Dist-Orc} to time Orc transitions eliminates the possibility of Zeno behaviors, which are a well-known artifact of logical time. This implies that for the intended semantics to be preserved, and hence correctness of the analysis later in Section \ref{sec:formal-analysis}, the transitions internal to an Orc configuration must be completed before the next real-time clock tick arrives. In other words, a single clock tick should be long enough to accommodate the instantaneous transitions of an Orc configuration. The minimum length of a clock tick so that this property is satisfied is specific  to the Orc application and the machines used to run it. For example, for the distributed auction case study below, and using a 2.0GHz dual-core node with 4GB of memory, the clock tick can be made as short as 0.2 seconds. In general, deciding on a minimum size for a clock tick given an application is hard to anticipate and is typically accomplished through experimentation. Normally, for distributed Orc applications, it is enough to make sure that the application is designed so that a one-second clock tick is long enough for the application.

\subsection{\texttt{Dist-Auction}: A Distributed Implementation of \textsc{Auction}} \label{sec:auction-example}

To illustrate our distributed implementation, we describe a distributed implementation \texttt{Dist-Auction} of the online auction management application in Orc, \textsc{Auction}, which was introduced in Section \ref{sec:orc}. The distributed configuration of the auction application contains two expression configurations: one with the $\mathit{Posting}$ expression object, which is responsible for retrieving and posting items for sale by a given seller, and the other contains the $\mathit{Bidding}$ expression object for managing the bidding process. For instance, the initial local configuration for the $\mathit{Posting}$ expression object has the form:

\begin{small}
\begin{verbatim}
[ <> 
  < C : Clock | clk : c(0) > 
  createServerTcpSocket(socketManager, C, 54200, 10)
  < e(loc("10.0.0.2", 44200), 0) : Expr | 
    env: Posting s := s("postNext") > x > AUCTIONID("post",x) >> rtimer(1) >> 
                      Posting(s), 
    exp: Posting(SELLERID), ... >  ... ]
\end{verbatim}
\end{small}

\noindent where \texttt{SELLERID} and \texttt{AUCTIONID} are object identifiers for the $\mathit{Seller}$ and $\mathit{Auction}$ sites, respectively. The $\mathit{Posting}$ expression declaration is stored in the environment attribute \texttt{env} of the expression object, while the attribute \texttt{exp} keeps the actual expression to be evaluated. The configuration also includes objects for internal (fundamental) sites, such as $\mathit{if}$ and $\mathit{let}$, which are omitted here for brevity.

In addition to the $\mathit{Posting}$ and $\mathit{Bidding}$ expression configurations, there are four site object configurations in the distributed configuration of \texttt{Dist-Auction}, one configuration for each of the sites assumed by \textsc{Auction}, namely $\mathit{Seller}$, $\mathit{Bidders}$, $\mathit{MaxBid}$, and $\mathit{Auction}$. For example, the initial local configuration for a  $\mathit{Seller}$ site with two items for auction (identified by numbers \texttt{1910} and \texttt{1720}) may have the form:

\begin{small}
\begin{verbatim}
[ <> 
  < C : Clock | clk : c(0) > 
  createServerTcpSocket(socketManager, C, 54800, 10)
  < s(loc("10.0.0.4", 44800), 0) : Site | 
      name : 'seller, state : (item(1910, 5, 500), item(1720, 7, 700)) , ... >
  createServerTcpSocket(socketManager, s(loc("10.0.0.4", 44800), 0), 44800, 10)]
\end{verbatim}
\end{small}

\noindent The site attempts to create two server sockets: one for listening to expression object requests and the other for listening to the local ticker object.

\begin{figure} 	
	\centering
	\includegraphics[width=0.90\textwidth]{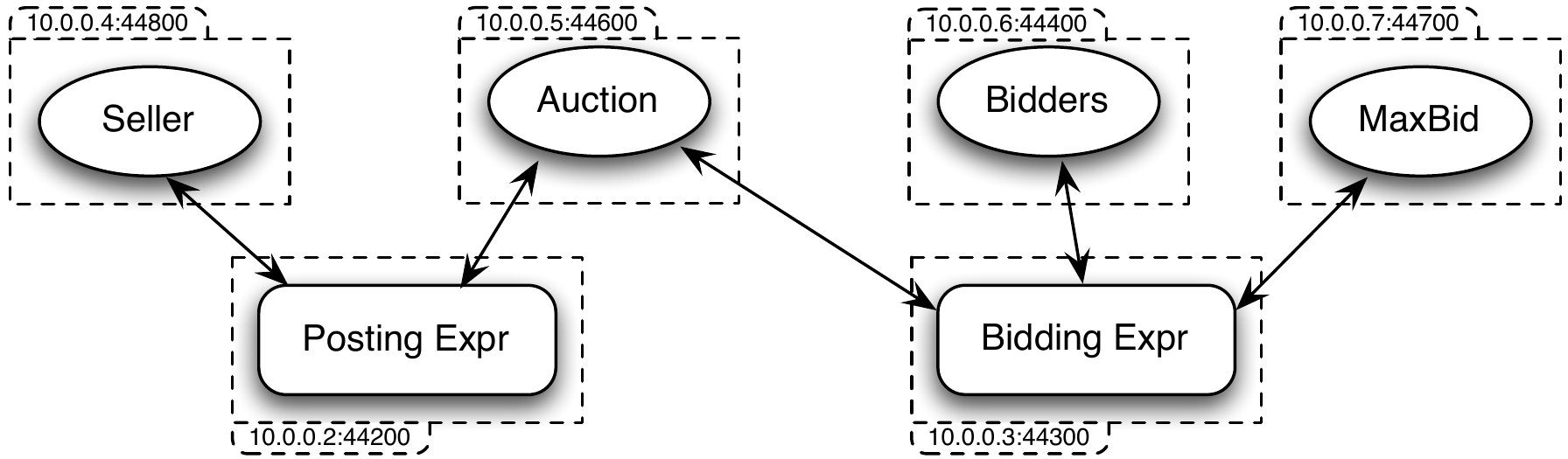}
	\caption{The deployment architecture of the \texttt{Dist-Auction} Orc program}
    \label{fig:Dist-Auction-arch}
\end{figure}

Each local configuration in \texttt{Dist-Auction} may run on a different node in a communication network. The diagram in Figure \ref{fig:Dist-Auction-arch} depicts a physical deployment of \texttt{Dist-Auction}, with bidirectional arrows representing communication patterns. A physical deployment can be conveniently achieved using an appropriate shell script to run Maude, load the \textsc{Dist-Orc} module and \texttt{Dist-Auction}, and execute the external rewrite command \texttt{erew}. For example, with \texttt{initAuction} an operator that creates the initial state of the Auction site configuration, the following command executes the Auction site:

\begin{small}
\begin{verbatim}
echo "erew initAuction ." | maude orc-distributed.maude auction-manager.maude 
\end{verbatim}
\end{small}

\noindent with the following sample output, generated by the \texttt{print} attribute of Maude statements (with auction items \texttt{1910} and \texttt{1720}):

\begin{tabular}{@{}p{.62\textwidth} p{.3\textwidth}@{}}
\begin{scriptsize}
\begin{verbatim}
00 erewrite in DIST-AUCTION : initAuction .
01 Site "10.0.0.5":44600 0 initializing... Site is ready.
02 Clock server socket created. 
03 Awaiting connection from ticker ...
04 Ticker connected.
05 Received "post"
06 Item 1910 posted
07 Received "getNext"
08 Bidding to start for 1910
09 Tick! ...  (5 time ticks)
10 Received "won"
\end{verbatim}
\end{scriptsize}  
&
\begin{scriptsize}
\begin{verbatim}
11 Item 1910 won by Bidder 3
12 Received "getNext"
13 Tick!
14 Received "post"
15 Item 1720 posted
16 Bidding to start for 1720
17 Tick! ...  (6 time ticks)
18 Received "won"
19 Item 1720 won by Bidder 3
20 Received "getNext"
21 ...
\end{verbatim}
\end{scriptsize}
\end{tabular}

In this run, the auction site receives a \texttt{post} request from the $\mathit{Posting}$ expression object and posts item \texttt{1910}. Meanwhile, a request for the next item to be auctioned is received from  the $\mathit{Bidding}$ expression object. The auction site then publishes the item details back to the $\mathit{Bidding}$ expression, which takes care of orchestrating the bidding process for this item. After five time units (the duration of the auction on item \texttt{1910}), Bidder 3 is announced as the winner and a similar process is repeated for the second item \texttt{1720}.

\section{Formal Analysis of Distributed Orc Programs} \label{sec:formal-analysis}

The implementation described above is very useful in prototyping and deploying Orc programs as it allows observing actual possible behaviors in realistic environments. However, the implementation technique outlined above does not result in a language specification that is immediately amenable to more rigorous formal analysis such as reachability analysis and model-checking. This is fundamentally due to the use of TCP sockets and ticker objects, which are outside the scope of the Maude formal analysis tools. While support for sockets is built into Maude, sockets do not have a logical representation that can be subjected to formal analysis. Furthermore, the ticker objects, being written in another general-purpose language with access to the system's real, wall clock time, introduce yet another obstacle in achieving a formally analyzable specification. Our solution to this problem, which we explain in this section, is to define rewriting logic specifications in Real-Time Maude for both Maude sockets and externally defined tickers, so that the distributed implementation can be turned, with minimal effort, into a formally analyzable specification.

\subsection{Formal Specification of TCP Sockets}

Maude's TCP sockets can be formally specified by defining abstractions of Maude instances, sockets, and their behaviors. We develop a rewriting specification $\mathcal{R}_{Socket}$ of sockets, which is based on previous work in~\cite{DuranRV07,RiescoV07}. 
The specification abstracts Maude instances with objects of the class $\mathit{Process}$, and server and client sockets with objects of $\mathit{ServerSocket}$ and $\mathit{Socket}$ classes, respectively, which  mediate communication between processes. A process object has the form $\langle \mathit{PID} : Process \;|\; \mathit{sys}: S \rangle$, with $\mathit{PID}$ an object identifier and $S$ an encapsulated local configuration. A client socket object of the form 
$\langle \mathit{SID} : Socket \;|\; endpoints : [\mathit{PID}_{a}, \mathit{PID}_{b}] \rangle$
 abstracts a bidirectional client TCP socket set up between processes $\mathit{PID}_{a}$ and $\mathit{PID}_{b}$, while a server socket object simply maintains the server's address and port: $\langle \mathit{SID} : \mathit{ServerSocket} \;|\; \mathit{address} : A, \mathit{port} : N \rangle$.
Server socket objects are created using a \emph{Manager object}, $\langle \mathit{socketManager} : \mathit{Manager} \;|\; \mathit{counter} : N \rangle$, abstracting Maude's socket manager and maintaining a counter for generating fresh socket object identifiers. Finally, messages in $\mathcal{R}_{Socket}$ have formats that are almost identical to those used by Maude sockets.

Different useful abstractions of socket behaviors can be defined. For the formal model of \textsc{Dist-Orc}, we choose an abstraction level that captures most interesting behaviors and yet can be efficiently analyzed. The abstraction essentially considers potential client socket creation errors and a somewhat limited form of communication delays and failures. This design choice abstracts away uninteresting behaviors, such as server socket creation problems, and approximates actual messaging problems, such as unavailable or unreachable servers, and unreliable networks. The main features of $\mathcal{R}_{Socket}$ are explained below.

Server socket creation is straightforward, and is modeled with the following rewrite rule:

\begin{small}
\begin{verbatim}
rl [CreateServerTcpSocket] : 
  < PID : Process | sys : [createServerTcpSocket(socketManager, O, PT) CONF] >
  < socketManager : Manager | counter : N >
  => < PID : Process | sys : [createdSocket(O, socketManager, server(N)) CONF] >
     < socketManager : Manager | counter : (N + 1) >
     < server(N) : ServerSocket | address : "localhost", port : PT > .
\end{verbatim}
\end{small}

\noindent The rule creates a server socket object, identified by \texttt{server(N)}, with an arbitrary address and a given port, and transforms the socket creation request message into an appropriate response. 

When an Orc object within a process attempts to create a client socket to a server by issuing the message \texttt{createClientTcpSocket(socketManager, O', SR, PT)}, two different transitions are possible depending on whether the client socket creation is successful or not. The success case is modeled by the following rule:

\begin{small}
\begin{verbatim}
rl [CreateClientSocketSuccess] : 
  < PID : Process | sys : [acceptClient(server(N), O) CONF] >
  < PID' : Process | sys : [createClientTcpSocket(socketManager, O', SR, PT) CONF' ] >
  < socketManager : Manager | counter : M >
  < server(N) : ServerSocket | address : SR, port : PT > 
  => < PID : Process | sys : [acceptedClient(O, server(N), SR, socket(M)) CONF] >
     < PID' : Process | sys : [createdSocket(O', socketManager, socket(M)) CONF'] >
     < socketManager : Manager | counter : (M + 1) > 
     < server(N) : ServerSocket | address : SR, port : PT >
     < socket(M) : Socket | endpoints : [PID : PID'] > .
\end{verbatim}
\end{small}

\noindent In this rule, the server is in a state accepting incoming connections from clients, specified by matching a server at address and port \texttt{SR:PT} that is accepting connections using the message \texttt{acceptClient(...)}. The rule also creates a socket object \texttt{socket(M)} that will mediate communication between the client and the server. 

Client socket creation may also fail, representing situations where the server is unreachable or unavailable. This case is modeled by the following rule, where the client process gets the \texttt{socketError(O', socketManager, "")} message from the socket manager, and no new socket object is created.  
 
\begin{small}
\begin{verbatim}
rl [CreateClientSocketFail] : 
  < PID : Process | sys : [acceptClient(server(N), O) CONF] >
  < PID' : Process | sys : [createClientTcpSocket(socketManager, O', SR, PT) CONF' ] >
  < socketManager : Manager | counter : M >
  < server(N) : ServerSocket | address : SR, port : PT >
  => < PID : Process | sys : [acceptClient(server(N), O) CONF] >
     < PID' : Process | sys : [socketError(O', socketManager, "") CONF'] >
     < socketManager : Manager | counter : M >
     < server(N) : ServerSocket | address : SR, port : PT >  . 
\end{verbatim}
\end{small}

Once a socket is successfully created, a connection through this socket is established, and bidirectional message exchanges may take place using \texttt{send(...)} and \texttt{receive(...)} messages.

\begin{small}
\begin{verbatim}
crl [exchange] : 
 < PID : Process | sys : [send(SOCKET, O, C) CONF] >
 < PID' : Process | sys : [receive(SOCKET, O') CONF'] >
 < SOCKET : Socket | endpoints : [PID : PID'] > 
 < DID : Delays | ds : DS > 
 => < PID : Process | sys : [sent(O, SOCKET) CONF] >
    < PID' : Process | sys : [received(O', SOCKET, C, R) CONF'] >
    < SOCKET : Socket | endpoints : [PID : PID'] > 
    < DID : Delays | ds : DS >
 if DS' R DS'' := DS .
\end{verbatim}
\end{small}

\noindent The \texttt{send} and \texttt{receive} messages are respectively transformed into a \texttt{sent(O, SOCKET)} message, acknowledging the send action to the sender, and a \texttt{received(O', SOCKET, C, R)} message, signaling \emph{delayed} delivery of the sent message to the receiver \emph{in \texttt{R}  time units}. 
The delay value for a transmitted message is non-deterministically extracted using a matching equation in the condition from a finite, non-empty set of delays \texttt{DS} maintained by a special object \texttt{DID}. 
A delay set for a given distributed Orc application can be specified as part of its initial state. Different behaviors may result by giving different delay sets. Two special cases of interest are: (1) $\mathit{DS} = \{ 0 \}$, in which case messages are assumed to experience no delays, and (2) $\infty \in  \mathit{DS}$, which represents the case of a lossy communication channel. As we will see in Section \ref{sec:global-logical-time}, the real-time semantics of the model will eventually make such delayed messages available to the receiver for processing. 

Finally, closing a socket is straightforwardly modeled by the following equation:

\begin{small}
\begin{verbatim}
eq [close] : 
  < PID : Process | sys : [closeSocket(SOCKET, O) CONF] >
  < PID' : Process | sys : [receive(SOCKET, O') CONF'] >
  < SOCKET : Socket | endpoints : [PID : PID'] > 
  = < PID : Process | sys : [closedSocket(O, socketManager, "") CONF] >
    < PID' : Process | sys : [closedSocket(O', socketManager, "") CONF'] > .
\end{verbatim}
\end{small}

\noindent The equation drops the closed socket, and issues the \texttt{closedSocket(...)} message to its endpoints.

\subsection{Global Logical Time} \label{sec:global-logical-time}

Time and its effects on the distributed Orc configuration are formally specified using the standard and general technique of capturing logical time in real-time rewrite theories~\cite{RTTheories05}, and facilitated by Real-Time Maude~\cite{RTMManual07}. 
Essentially, the time domain is represented by a sort $\mathit{TimeInf}$ (time with infinity), and a \emph{global tick} rewrite rule is used to synchronously advance time and propagate its effects across the \emph{encapsulated global configuration}, a term of the sort $\mathit{GlobalSystem}$, of the form $\{ \mathit{Conf} \}$, where $\mathit{Conf}$ is the Orc configuration consisting of all process and socket objects. 
Furthermore, the tick rule, which plays the role of the ticker objects in the distributed implementation, is defined globally as follows (with \texttt{R'} a variable ranging over the positive rational numbers): 

\begin{small}
\begin{verbatim}
crl [tick] : 
  {CONF} => {delta(CONF, R')} in time R'
    if eagerEnabled({CONF}) =/= true /\ R' <= mte(CONF) [nonexec] .
\end{verbatim}
\end{small}

\noindent The tick rule computes on the global Orc configuration the function \texttt{delta}, which advances time for all clock objects and updates time delays in all site calls and returns in the configuration. For example, clocks and delayed external messages are updated, respectively, by the following two equations (\texttt{plus} and \texttt{monus} define addition and subtraction on time domains):

\begin{small}
\begin{verbatim}
eq delta(< O : Clock | clk : c(R) > CF, R') 
   = < O : Clock | clk : c(R plus R') > delta(CF, R') . 
eq delta(received(O, O', C, R) CF, R') = received(O, O', C, R monus R') delta(CF, R') .
\end{verbatim}
\end{small}

The tick rule above is not immediately executable (which is indicated by the \texttt{[nonexec]} attribute), as it introduces a new variable \texttt{R'} representing the amount of time elapse on its right hand side. A strategy for sampling time needs to be specified for the rule to be executable. We assume a general \emph{maximal} strategy that in each tick advances time by the \emph{maximum time elapse}, which is defined by the function \texttt{mte} as the minimum delay across all site call messages and returns in the global Orc configuration. This ensures that time is advanced as much as possible in every tick but only enough to be able to capture all events of interest. To properly capture the synchronous semantics of Orc~\cite{AlTurkiM07PPDP,AlTurkiM07WWV}, the tick rule is also made conditional on the fact that no other instantaneous transition is possible. This is precisely captured by the \texttt{eagerEnabled} predicate, which  is true on configurations with enabled internal transitions or pending message exchanges. This imposes a precedence of rule application, where time ticks have the lowest priority among all transitions. 

It is important to note that the abstraction of time and how it affects the global Orc configuration as specified by the tick rule is consistent with the real-time distributed implementation \textsc{Dist-Orc} in that, in \textsc{Dist-Orc}, we assumed that the granularity of a single time tick in real-time is always large enough for instantaneous transitions within a configuration to complete. Furthermore, the tick rule \emph{synchronously} updates all clock objects in all processes. This defines yet another abstraction over \textsc{Dist-Orc}, where individual clocks are not necessarily synchronized. However, since clock synchronization is not required for \textsc{Dist-Orc}, as was discussed in Section \ref{sec:timed-implementation}, the abstraction considers only those behaviors in \textsc{Dist-Orc} that make sense under these assumptions about time.

\subsection{Formal Analysis of \texttt{Dist-Auction}}

The formal specification of sockets and logical time provides a formal model of \textsc{Dist-Orc} that can be used to verify properties about distributed applications in Orc. To illustrate this formal verification capability, we use Real-Time Maude to formally analyze the distributed implementation  \texttt{Dist-Auction} of the auction case study. In particular, we perform time-bounded linear temporal logic model checking with commands of the form \texttt{(mc \emph{term} |=t \emph{formula} in time <= \emph{timeLimit} .)}, and timed search using \texttt{(find earliest \emph{term} =>* \emph{pattern} such that \emph{condition} .)}, which finds a state reachable within the shortest possible time that matches the given pattern and satisfies the given condition.
Verification is applied on a \emph{closed} system specification that includes definitions of all required sites (servers) and expressions (clients) in the \textsc{Auction} application.

In our analysis, we assume a single seller site with two items for sale, labeled \texttt{1910} and \texttt{1720}, and offered for auction for $5$ and $7$ time units, respectively. The function $\texttt{initial(DS)}$ constructs an initial state for \texttt{Dist-Auction} in which the set of possible message transmission delays is~$\texttt{DS}$, which, in the analysis examples below, is the singleton set $\{ 0.1 \}$, unless otherwise indicated. The atomic predicates used are: 
\begin{enumerate}
\item $\mathit{commError}$, which is true in states with communication errors:

\begin{small}
\begin{verbatim}
op commError : -> Prop .
eq {< PID: Process | sys: [socketError(O, O', S) CF] > CF'} |= commError = true .
\end{verbatim}
\end{small}

\item $\mathit{sold}(id)$, which is true in states where the item $id$ has been sold: 

\begin{small}
\begin{verbatim}
op sold : Nat -> Prop .
eq {< PID : Process | 
        sys : [< O : XSite | name : 'auction, 
                             state : won(winner(N, id, M), WN) OST > CF] > CF'} 
    |= sold(id) = true .
\end{verbatim}
\end{small}

\noindent where the term \texttt{winner(N, id, M)} matches a winning bid \texttt{M} on item \texttt{id} by the \texttt{N}th bidder.

\item $\mathit{hasBid}(id)$, which is true when the item $id$ has been bid on:

\begin{small}
\begin{verbatim}
op hasBid : Nat -> Prop .
eq {< PID : Process | 
        sys : [< O : XSite | name : 'bidders, 
                             state : bidders(b(N, [id, M] IBS) BS) OL > CF] > CF'} 
    |= hasBid(id) = true .
\end{verbatim}
\end{small}

\noindent where the term \texttt{b(N, [id, M] IBS)} matches a bid \texttt{M} on item \texttt{id} by the \texttt{N}th bidder.

\item $\mathit{conflict}(id)$, which is true when item $id$ has two different winners:

\begin{small}
\begin{verbatim}
op conflict : Nat -> Prop .
eq {< PID : Process | 
        sys : [< O : XSite | name : 'auction , 
                             state : won(winner(N, id, M), winner(N', id, M'), WN) 
                                     OST > CF] > CF'} 
   |= conflict(id) = true .
\end{verbatim}
\end{small}

\end{enumerate}


A property that is typically required in an auction management system is that an item with at least one bid is eventually sold: $\Box \bigwedge_{i} (\mathit{hasbid}(id_{i}) \to \Diamond \mathit{sold}(id_{i}))$. This can be shown to be guaranteed by \texttt{Dist-Auction} in the absence of communication problems and excessively large delays. The property itself is specified in Real-Time Maude as the following formula \texttt{commitAllNoErrors} (with \texttt{\~} denoting the LTL negation operator) : 

\begin{small}
\begin{verbatim}
op commit : Nat -> Formula .
eq commit(id) = hasBid(id) -> <> sold(id) .
op commitAllNoErrors : -> Formula .
eq commitAllNoErrors = ([] ~ commError) -> [] (commit(1910) /\ commit(1720)) .
\end{verbatim}
\end{small}

The property is then verified with the time-bounded model checking command:  

\begin{small}
\begin{verbatim}
Maude> (mc initial(1/10) |=t commitAllNoErrors in time <= 15 .)
rewrites: 7052663 in 14413ms cpu (14420ms real) (489317 rewrites/second) ...
Result Bool : true
\end{verbatim}
\end{small}

\noindent The property is satisfied with a communication delay of $0.1$ time units. In fact, the property is satisfied when communication delays are bounded by $0.25$ time units. This is because the timeout value for collecting bids in a single bidding round in the \emph{TimeoutRound} expression is $1.0$, while a delay of $0.25$ translates into a cumulative round trip delay of $1.0$ for its two sequential site calls, which may result in an uncommitted bid. This can be verified by the resulting counterexample when executing the command above but with \texttt{initial(1/4)}. 

Another property an auction management system must guarantee is that every item sold has a unique winner: $\Box \bigwedge_{i} \neg \mathit{conflict(id_{i})}$. This property can be shown satisfiable in \texttt{Dist-Auction} regardless of communication errors. The property is specified in Real-Time Maude as the formula \texttt{uniqueWinnerAll}:

\begin{small}
\begin{verbatim}
op uniqueWinner : Nat -> Formula .
eq uniqueWinner(id) = ~ conflict(id) .
op uniqueWinnerAll : -> Formula .
eq uniqueWinnerAll = [] (uniqueWinner(1910) /\ uniqueWinner(1720)) .
\end{verbatim}
\end{small}

The property is verified by the following command :

\begin{small}
\begin{verbatim}
Maude> (mc initial(1/10) |=t uniqueWinnerAll in time <= 15 .)
rewrites: 8613539 in 19627ms cpu (19800ms real) (438843 rewrites/second) ...
Result Bool : true
\end{verbatim}
\end{small}

Finally, given a delay of $0.1$, one can verify that the first item cannot be won before $5.5$ time units have passed using the following command: 

\begin{small}
\begin{verbatim}
Maude> (find earliest initial(1/10) =>* {C:Configuration} 
          such that {C:Configuration} |= sold(1910) .)
rewrites: 268287407 in 1525921ms cpu (1544117ms real) (175819 rewrites/second) ...
Result:	{< did : Delays | ds : 1/10 > ... } in time 11/2
\end{verbatim}
\end{small}


\section{Conclusion and Future Work} \label{sec:concl}


We have presented \textsc{Dist-Orc}, a rewriting-based, real-time, distributed implementation of the Orc language allowing different Orc expressions at different locations to interact by asynchronous message passing with different sites in an object-based manner.  We have also shown how a \textsc{Dist-Orc} real-time implementation can be easily obtained from a rewriting logic semantic definition of Orc in a correct-by-construction way using Maude sockets and ticker objects.  And we finally demonstrated with an auction example that Orc applications running in \textsc{Dist-Orc} can still be formally analyzed by model checking once we model the distributed infrastructure at a reasonable level of abstraction.

Much work remains ahead. Besides developing a broader class of examples and
optimizing the present prototype, three interesting future directions are: (i) developing a transformation method based on the techniques presented here that can automatically synthesize a real-time, distributed implementation from the formal semantics; (ii) making \textsc{Dist-Orc} more user-friendly, by providing a user interface
for interacting with \textsc{Dist-Orc}-based web orchestration applications; and (iii) endowing
\textsc{Dist-Orc} with a \emph{security infrastructure}, and formally verifying the security of certain
types of web orchestration services that use such an infrastructure against general classes of
attacks.

\subsubsection*{Acknowledgements}
We thank Jayadev Misra, William Cook, David Kitchin, Adrian Quark and Andrew Matsuoka for very fruitful discussions on Orc and its Java implementation, and the anonymous reviewers and Peter \"{O}lveczky for helpful comments on an earlier version of the paper. This work has been partially supported by NSF grant NSF CNS 07-16638, King Fahd University of Petroleum and Minerals, and King Abdullah University of Science and Technology.

%
%
\bibliographystyle{splncs}
\bibliography{ds-rtrts10}

\end{document}